\magnification=1200
\settabs 18 \columns

\topinsert \vskip .50 in
\endinsert

\def\s{\smallskip}

\def\sqr#1#2{{\vcenter{\vbox{\hrule height.#2pt
 \hbox{\vrule width.#2pt height#1pt \kern#1pt
 \vrule width.#2pt} \hrule height.#2pt}}}}

\def\operp{\hbox{${\kern+.25em{\bigcirc}
\kern-.85em\bot\kern+.85em\kern-.25em}$}}

\def\lsim{\;\raise0.3ex\hbox{$<$\kern-0.75em\raise-1.1ex\hbox{$\sim$}}\;}
\def\gsim{\;\raise0.3ex\hbox{$>$\kern-0.75em\raise-1.1ex\hbox{$\sim$}}\;}
\def\no{\noindent}

\def\ce{\centerline}
\def\ve{\vfill\eject}
\def\rdots{\mathinner{\mkern1mu\raise1pt\vbox{\kern7pt\hbox{.}}\mkern2mu
 \raise4pt\hbox{.}\mkern2mu\raise7pt\hbox{.}\mkern1mu}}

\def\e e{$e^+ e^-$ }

\input epsf

\ce {GENERAL RELATIVISTIC SOLITONS (II)}
\vskip.5cm
\ce{A. C. Cadavid}
\s

\ce{\it Department of Physics and Astronomy}
\ce{\it California State University, Northridge, CA 91330}
\s

\ce{R. J. Finkelstein}
\s

\ce{\it Department of Physics and Astronomy}
\ce{\it University of California, Los Angeles, CA  90095-1547}
\vskip 1cm

\no {\bf Abstract.} We investigate the possible existence of
non-topological solitons in string-like theories, or in other
completions of Einstein theory, by examining a simple
extension of standard theory that describes a non-linear scalar
field interacting with the Einstein, Maxwell and Weyl
(dilaton) fields. The Einstein and Maxwell couplings are
standard while the dilatonic coupling is taken to agree
with string models. The non-linear
scalar potential is quite general. It is found to be impossible
to satisfy the dilatonic boundary conditions. Excluding the 
dilaton field we find a variety of solitonic structures
differing in ways that depend on the non-linear potential.
In general the excited states exhibit a discrete mass spectrum.
At large distances the gravitational field approaches the 
Reissner-Nordstrom solution.

\ve
\ce {I. INTRODUCTION}
\vskip.5cm

Since the ``black hole solitons" of string and supergravity theories carry
horizons and central singularities, they differ from regular solitons
which are singularity-free.  It is natural to ask if these generally
covariant theories also admit regular non-topological solitons, similar
to the field structures that exist at the level of special relativity,$^{1,2}$
since it is known that non-topological solitons also exist at the
general relativistic level.$^{3,4,5}$
It is possible, however, that there
are generic features of supergravity and of other fundamental completions
of the Einstein theory that preclude the existence of everywhere regular
solitons that are non-topological. In particular, this role may be played by the dilaton field, which
arises as a consequence of Weyl rescaling, couples to the antisymmetric
tensor fields, and replaces the numerical coupling constant.

As these theories are derived by compactification from higher dimensional
formulations, they characteristically contain multiplets of nonlinear scalar
fields that are coupled to multiplets of vector and axial vector fields, and
higher tensor fields as well.  They are therefore difficult to test for
everywhere regular solutions since their field structures, although
mitigated by large symmetry groups, are very complex.  Nevertheless it may be
interesting to examine model theories obtained by coupling 
the dilaton field and a nonlinear scalar
multiplet to the gravitational and electromagnetic fields, both for their
intrinsic interest and also for their similarity to more fundamental
theories.  We shall consider a model theory with the following action which
exhibits some of these features.
$$
S = \int d^4x\sqrt{-g}(R-KL) \eqno(1.1)
$$
$$ \eqalign{
L = {1\over 2} &g^{\mu\lambda} \nabla_\mu\psi_0\nabla_\lambda\psi_0 -
{1\over 2} e^{\epsilon\psi_0} F_{\mu\lambda}F^{\mu\lambda} \cr
&+ \sum^N_{i=1} g^{\mu\lambda}\nabla_\mu\psi_i(\nabla_\lambda\psi_i)^* -
V(\sum^N_{i=1} \psi_i\psi^*_i) \cr} \eqno(1.2)
$$ 
\no where the neutral scalar $\psi_0$ is coupled to $F_{\mu\lambda}$ as a
dilaton, where $\epsilon > 0$, 
and where the charged scalars are coupled in the standard way to both
the electromagnetic and gravitational fields by the following covariant
derivative
$$
\nabla_\mu = \partial_\mu + {ie\over c\hbar} A_\mu + \Gamma_\mu~. \eqno(1.3)
$$
\no Here $\Gamma_\mu$ is the gravitational part of the covariant
derivative and $V$ is the nonlinear scalar interaction.  It would be possible
to include magnetic couplings in (1.2) but we shall consider only the case
in which magnetic field and magnetic charge vanish.  We have not included any
self-coupling of the dilaton field.

When $\psi_0$ is required to be constant, we have a theory with standard
couplings.

One may think of this model theory as interpolating between standard
physics \break
$\psi_0$ = constant, and new physics 
associated with a Kaluza-Klein theory.  It is not, however, intended as a truncation of a
supergravity theory.
\ve

\ce{II. THE SOLITON PROBLEM}
\vskip.5cm

In the action (1.1) there are the following coupled fields: the metric
$g_{\mu\lambda}$, the vector potential $A_\mu$, and the scalar
fields $\psi_i$.  The equations of motion are
$$
\eqalignno{{\delta S\over\delta g_{\mu\lambda}} &= 0 & (2.1) \cr
{\delta S\over \delta A_\mu} &= 0 & (2.2) \cr
{\delta S\over \delta\psi_i} &= 0~~, \quad i=0,\ldots,N & (2.3) \cr}
$$

We require everywhere regular solutions of these equations subject to
solitonic boundary conditions.  Therefore no central singularities or
horizons are permitted.  These solitonic solutions concentrate 
energy-momentum and charge-current within lumps of field.  In general,
the mass spectrum of these particle-like lumps may be discrete.

In this theory both the nonlinear scalar fields and the Maxwell field
contribute to the source of the gravitational field while the complex
scalars are the source of the electromagnetic field.  To avoid irrelevant
complications, it is assumed that the charged scalar fields are spherically
symmetric and harmonically time-dependent, while the dilaton is also
spherically symmetric but time-independent:
$$
\eqalignno{\psi_0(x) &= R_0(r) & (2.4) \cr
\psi_i(x) &= e^{i\omega t} R_i(r)~, \quad i=1,\ldots N & (2.5) \cr}
$$
\no We shall need to consider only one member of the $SU(N)$ scalar multiplet.

Since we are dropping the magnetic fields, the vector potential also vanishes
and the scalar potential is spherically symmetric.  Thus the electric
and gravitational fields should resemble the Reissner-Nordstrom solution
except near the origin where the mass and charge are spread out and all fields
remain finite.
\ve

\ce{III.  THE FIELD EQUATIONS}
\vskip.5cm

The field equations implied by (1.2) and (1.3) may be expressed as follows:
\item{(a)} The gravitational equations
$$
R_{\mu\lambda} = K\Theta_{\mu\lambda} \quad
K = -{8\pi k\over c^2} \eqno(3.1)
$$
\item{} where $k$ is Newton's constant.  Here
$$
\eqalignno{\Theta_{\mu\lambda} &= \theta_{\mu\lambda} - {1\over 2}
\theta g_{\mu\lambda} & (3.2) \cr
\theta_{\mu\lambda} &= {\partial L\over\partial g^{\mu\lambda}} -
{1\over 2} g_{\mu\lambda}L~. & (3.3) \cr}
$$

\item{(b)} The scalar equations
$$
\eqalignno{g^{\mu\lambda}\nabla_\mu\nabla_\lambda\psi_0 + {\epsilon\over 2}
e^{\epsilon\psi_0} F_{\mu\lambda}F^{\mu\lambda} &= 0 & (3.4) \cr
g^{\mu\lambda}\nabla_\mu\nabla_\lambda\psi_i +
{\partial V\over\partial\psi_i^*} &= 0 & (3.5) \cr}
$$
\item{} The operator $\nabla_\mu$ contains the $A_\mu$ term in (3.5)
but not in (3.4).

\item{(c)} The electromagnetic equations
$$
{1\over\sqrt{-g}} \partial_\lambda
(\sqrt{-g}~e^{\epsilon\psi_0}F^{\mu\lambda}) =
{ie\over 4\hbar c} \sum [\psi_i(\nabla^\mu\psi_i)^* -
\psi_i^*(\nabla^\mu\psi_i)]~. \eqno(3.6)
$$

\no The usual equation for the conserved current follows from (3.6), namely
$$
\partial_\mu\bigl\{\sqrt{-g}\sum^N_{i=1}
[\psi_i(\nabla^\mu\psi_i)^*-\psi_i^*(\nabla^\mu\psi_i)]\bigr\} = 0~.
\eqno(3.7)
$$
\no To simplify the general equations (3.1), choose the line element
$$
ds^2 = e^{\nu(r)}c^2dt^2-e^{\lambda(r)}dr^2
-r^2(d\theta^2+\sin^2\theta d\varphi^2) \eqno(3.8)
$$
\no and in (3.4) and (3.5) choose
$$
F_{\alpha\beta} = \left(\matrix{0 & -E & 0 & 0 \cr
E & 0 & 0 & 0 \cr 0 & 0 & 0 & 0 \cr 0 & 0 & 0 & 0 \cr} \right) \eqno(3.9)
$$
\no where
$$
E = -{\partial\varphi\over\partial r} = -\varphi^\prime \eqno(3.10)
$$
\no and $\varphi$ is $A_0$, the scalar potential of the electromagnetic
field.  The complete set of gravitational equations is now
$$
\eqalignno{-&{\nu^{\prime\prime}\over 2} + {\nu^\prime\lambda^\prime\over 4}
-{(\nu^\prime)^2\over 4}-{\nu^\prime\over r} =
K\biggl[e^{\lambda-\nu}\biggl({\omega\over c}+{e\over\hbar c}\varphi\biggr)^2
R^2 - {e^\lambda\over 2} V(R^2)+\Delta\biggr] & (3.11) \cr
& {\nu^{\prime\prime}\over 2}-{\nu^\prime\lambda^\prime\over 4}+
{(\nu^\prime)^2\over 4} - {\lambda^\prime\over r}= K
\biggl[{1\over 2}(R_0^\prime)^2 + (R^\prime)^2 + {1\over 2}
e^\lambda V(R^2)-\Delta\biggr] & (3.12) \cr
&(\nu^\prime-\lambda^\prime){r\over 2} = K
\biggl[{r^2\over 2} V(R^2) e^\lambda + r^2\Delta\biggr]
+ e^\lambda -1 ~. & (3.13) \cr}
$$
\no 
The off-diagonal equations are satisfied identically.

Here
$$
\eqalignno{\Delta &= e^{\epsilon R_0-\nu} (\varphi^\prime)^2/2c^2 & (3.14) \cr
R^2 &= \sum^N_1 R_i^2 ~. & (3.15) \cr}
$$
\no The equations for the dilaton and the charged scalars are:
$$
\eqalignno{&R_0^{\prime\prime} + {2\over r} R_0^\prime +
{\nu^\prime-\lambda^\prime\over 2} R_0^\prime + \epsilon
e^{\epsilon R_0-\nu} (\varphi^\prime)^2 = 0 & (3.16) \cr
&R_i^{\prime\prime} + {2\over r} R_i^\prime +
{\nu^\prime-\lambda^\prime\over 2} R^\prime_i +
\biggl({\omega\over c} + {e\over\hbar c}\varphi\biggr)^2
e^{\lambda-\nu} R_i-e^\lambda {\partial\over\partial R_i}
V(R^2) = 0~, \quad i=1\ldots N~. & (3.17) \cr}
$$
\no Finally the single equation for the electromagnetic scalar potential
is
$$
\varphi^{\prime\prime} + {2\over r} \varphi^\prime -
\biggl[{1\over 2}(\nu^\prime + \lambda^\prime)+\epsilon R_0^\prime\biggr]
\varphi^\prime = -{1\over 2} {e\over\hbar c}
\biggl({\omega\over c}+{e\over\hbar c}\varphi\biggr)
e^{-\epsilon R_0+\lambda} R^2~. \eqno(3.18)
$$
\no The boundary conditions may be chosen as follows:
$$
\eqalignno{R_i^\prime(0) &= 
\varphi^\prime(0) = 0 & (3.19) \cr
R_i(\infty) &= \lambda(\infty) = \nu(\infty) = \varphi(\infty) = 0 & (3.20) \cr
R_i^\prime(\infty) &= \lambda^\prime(\infty) = \nu^\prime(\infty) =
\varphi(\infty) = 0 & (3.21) \cr}
$$
\no where $i=0,\ldots,N$.  In addition we require
$$
V(\infty) = 0~. \eqno(3.22)
$$

Since the $R_i$ are scalars, the equations (3.19) are invariant conditions
which exclude cusps at the origin in these fields.

The corresponding restriction on $\varphi$ is
$$
\varphi^\prime(0) = E(0) = 0 \eqno(3.23)
$$
\no which is required since the origin is at the center of a spherically
symmetric charge distribution.  In invariant form (3.23) reads
$$
\left(F_{\mu\lambda}F^{\mu\lambda}\right)_{r=0} = 0~. \eqno(3.23)'
$$

In the rest of this paper we shall use the gauge where
$$
\eqalignno{R_i &= 0~, \qquad i=2,\ldots,N & (3.24) \cr
R_1 &= R ~. \cr}
$$
\no We shall also need the following relation
$$
-{(\nu^\prime+\lambda^\prime)\over r} =
K\biggl[(R^\prime)^2 + {1\over 2}(R_0^\prime)^2 +
\biggl({\omega\over c}+{e\over\hbar c}\varphi\biggr)^2
e^{\lambda-\nu} R^2/c^2\biggr] \eqno(3.25)
$$
\no obtained by adding (3.11) and (3.12).
It follows from the preceding equation and the boundary conditions
that
$$
\nu^\prime(0) + \lambda^\prime(0) = 0~.  \eqno(3.26)
$$
\no The invariant form of (3.26) may be expressed in terms of $g$, the
determinant of the metric tensor, as follows:
$$
\left( {g \over g_{flat}} \right)_{r=0}^\prime = 0 \eqno(3.27)
$$

\no One also notes that (3.13) implies
$$
\nu^\prime(0) -\lambda^\prime(0) =
\biggl({2\over r} (e^\lambda -1)\biggr)_{r=0} ~. \eqno(3.28)
$$
\no Here
$$
\lambda = \ell_0 + \ell_2 r^2 + \ldots \eqno(3.29)
$$
\no since $\lambda(r)$ and $\nu(r)$ are even by (3.11)-(3.13).  If
$$
e^{\lambda(0)} \not= 1 \eqno(3.30)
$$
\no then by (3.28)
$$
\nu^\prime(0) - \lambda^\prime(0) = \infty \eqno(3.31)
$$
\no and by (3.26) both $\nu(r)$ and $\lambda(r)$ 
would have infinite cusps at $r=0$.
If
$\lambda(0)$ vanishes, however, then by (3.28)
$$
\nu^\prime(0)-\lambda^\prime(0)=0~. \eqno(3.32)
$$
\no Therefore we adopt the following boundary conditions
$$
\eqalignno{\lambda(0) &= 0  & (3.33) \cr
\lambda^\prime(0) &= \nu^\prime(0) = 0 ~. & (3.34) \cr}
$$
\no These boundary conditions are compatible with (3.26) and (3.32) as
well as with (3.11), (3.12) and (3.13).  Finally by (3.25) we have
$$
\bigl[\nu+\lambda\bigr]^\infty_0 = -K \int^\infty_0
drr\biggl[(R^\prime)^2 + {1\over 2} (R^\prime_0)^2 +
\biggl({\omega\over c} + {e\over \hbar c} \varphi\biggr)^2
e^{\lambda-
\nu} {R^2\over c^2} \biggr] \eqno(3.35)
$$
\no and since the integrand is positive,
while $K$ is negative,
$$
\bigl[\nu + \lambda\bigr]^\infty_0 > 0 ~. \eqno(3.36)
$$
\no Then by (3.20)
$$
\nu(0) + \lambda(0) < 0~. \eqno(3.37)
$$
\no By (3.33)
$$
\nu(0) < 0~. \eqno(3.38)
$$

\ve

\ce {IV.  APPROXIMATE SOLUTIONS}
\vskip.5cm

The boundary conditions (3.19) and (3.20) imply
$$
\lim_{\scriptstyle r\to 0\atop\scriptstyle r\to\infty}
\Delta(r) = 0 \eqno(4.1)
$$
\no Therefore (3.11)-(3.13) all approach the corresponding equations
(except for the frequency shift $\omega + e\varphi/\hbar$) that hold in
the absence of a charged source.  These equations were previously studied.$^4$
Their solutions remain finite and have flat tangents at $r=0$.
\vskip.5cm

\no (a) The Electromagnetic Scalar Potential.

Let us consider (3.18) near $r=0$.  Guided by earlier results,$^4$
let us set
$$
\eqalignno{R_0 &= \hat a_0 + \hat a_2r^2 + \ldots & (4.2) \cr
R &= a_0 + a_2r^2 + \ldots  & (4.3) \cr
\lambda &= \ell_2 r^2 + \ell_4r^4 + \ldots & (4.4) \cr
\nu &= n_0 + n_2r^2 + n_4r^4 + \ldots & (4.5) \cr}
$$
\no Then
$$
R_0^\prime(0) = R^\prime(0) = \lambda^\prime(0) =
\nu^\prime(0) = 0 \eqno(4.6)
$$
\no and (3.18) near $r=0$ becomes
$$
\varphi^{\prime\prime} + {2\over r} \varphi^\prime \cong 
-{e\over 2\hbar c}\biggl({\omega\over c} + {e\varphi\over\hbar c}
\biggr) e^{-\epsilon\hat a_0} a_0^2 ~. \eqno(4.7)
$$
\no Set
$$
y = {\omega\over c} + {e\varphi\over \hbar c}~. \eqno(4.8)
$$
\no Now (4.7) becomes
$$
y^{\prime\prime} + {2\over r} y^\prime \cong -{1\over 2}
{e^2\over \hbar^2 c^2} e^{-\epsilon \hat a_0}a_0^2y~. \eqno(4.9)
$$
\no The solution of (4.9) is
$$
y = {\sin pr\over r} \eqno(4.10)
$$
\no where
$$
p^2 = {e^2\over 2\hbar^2c^2} e^{-\epsilon\hat a_0} a_0^2~. \eqno(4.11)
$$
\no Therefore
$$
\lim_{r\to 0}\varphi = {1\over \sqrt{2}} \exp\biggl[-\epsilon {\hat a_0
\over 2}\biggr]
a_0 - {\hbar\over e}\omega~. \eqno(4.12)
$$
\no Hence all fields including $\varphi(r)$ remain finite with vanishing
derivatives at $r=0$.  On the other hand at very large distances (3.18)
becomes
$$
\varphi^{\prime\prime} + {2\over r} \varphi^\prime = 0 \eqno(4.13)
$$
\no and
$$
\varphi = {Q\over r} \eqno(4.14)
$$
\no where $Q$ is the charge.
\vskip.5cm

\no (b) The Dilaton.

Let us next consider the equation of the dilaton (3.16).  Since all first
derivatives vanish at the origin, it follows from this equation that
$R^{\prime\prime}_0$ and $R^{\prime\prime\prime}_0$ also vanish at this
point.  The first non-vanishing derivative at the origin is
$$
R_0^{\rm IV}(0) = -2\epsilon e^{-\nu(0)+\epsilon R_0(0)}
(\varphi^{\prime\prime}(0))^2~. \eqno(4.15a)
$$
Let us first assume that $\epsilon>0$. Then $R_{0}(r)$ has a very flat
maximum at the origin, and therefore near the origin
$$
R^{\prime}_{0}(r) < 0 \eqno(4.15b)
$$
\no Equation (3.16) also implies that  $R_{0}(r)$ has no minimum anywhere
since
$$
R^{\prime \prime}_{0}(r) = - \epsilon e^{-\nu+\epsilon R_{0}} 
(\phi^{\prime})^{2} \leq 0 \eqno(4.16)
$$
\no wherever $R^{\prime}_{0}(r)$ vanishes.

To determine the behavior of $R_{0}(r)$ at very large $r$ one may anticipate
the result that $\lambda$ and $\nu$ both decrease as $1/r$. Then (3.16)
becomes approximately
$$
R^{\prime \prime}_{0}(r)+{2\over r} R^{\prime}_{0}(r)+
\epsilon (\phi^{\prime})^{2} \cong 0 \eqno(4.17)
$$
\no and if one sets 
$$
R_{0} = A r^{s} + B \eqno(4.18)
$$
$$
\phi= {Q\over r} \eqno(4.19)
$$
\no one finds by (4.17)
$$
R_{0}(r) = { -\epsilon Q^{2} \over 2 r^{2}} + B \eqno(4.20a)
$$
\no To satisfy the boundary condition at infinity set $B=0$.
By (4.20a) one has at large $r$
$$
R^{\prime}_{0}(r) = {\epsilon Q^{2} \over r^{3}} > 0 \eqno(4.20b)
$$

\no Now (4.15b) and (4.20b) are in conflict unless there is 
an intervening minimum.
But (4.16) does not allow a minimum anywhere, and so it is not
possible to satisfy the boundary conditions at both
$r=0$ and $r=\infty$. If we assume the opposite sign of
$\epsilon$, then there will be a minimum instead of a maximum
at the origin and all the inequalities will be reversed;
but there will remain the same incompatibility between the 
boundary conditions at the origin and at infinity. Since the 
presence of the dilaton field appearing in (1.2) therefore
precludes the existence of a non-topological soliton, we
shall set $\psi_{0}=0$ in (1.2) and delete $R_{0}$ from
all the equations following this paragraph. The resulting
action and equations are then standard, but the potential
remains general.

\vskip 0.5cm

\no (c) The Gravitational Equations.

Let us next examine the set (3.11)-(3.13) at $r=\infty$ subject to the
boundary conditions (3.20)-(3.21).  (We shall anticipate the result that
$R$ vanishes faster than $\nu$ and $\lambda$ as $r\to\infty$.)  Then by (3.25)
we have at large $r$ 
$$
\nu^\prime+\lambda^\prime \cong 0~. \eqno(4.21)
$$
\no By (3.20)
$$
\nu(\infty) + \lambda(\infty) = 0~. \eqno(4.22)
$$
\no Therefore at large distances
$$
\nu + \lambda \cong 0~. \eqno(4.23)
$$
\no Then by (3.13)
$$
-{\lambda^\prime\over r}\to K\biggl[{1\over 2} V(R^2) e^\lambda +
{e^{\lambda}\over 2c^2} (\varphi^\prime)^2\biggr] +
{e^\lambda-1\over r^2} \eqno(4.24)
$$
\no where we have also substituted (3.14) and (4.23). 

By the boundary conditions $V(R^2)$ vanishes at infinity.
We know that $\lambda$ vanishes as $1/r$ 
and it will be seen that $R$ vanishes as a spherical Bessel
function.  Then (4.24) becomes
$$
-{\lambda^\prime\over r}\cong {K\over 2c^2} (\varphi^\prime)^2
e^{\lambda} + {e^\lambda-1\over r^2}~. \eqno(4.25)
$$
\no To second order in $1\over r$, let us set
$$
\eqalignno{\lambda &\cong {2m\over r} + {\ell_2\over r^2} & (4.26) \cr
\varphi &\cong {Q\over r} + {q_2\over r^2} & (4.27) \cr}  
$$
\no Then by examining terms of order ${1\over r^4}$ one sees that
$$
\ell_2 = 2m^2 + {KQ^2\over 2c^2}~. \eqno(4.28)
$$
\no Note that
$$
e^\nu = (e^{-\lambda}) = 1 - {2m\over r} + {2m^2-\ell_2\over r^2}~. \eqno(4.29)
$$
\no Then by (4.28)
$$
e^\nu = 1 - {2m\over r} - {KQ^2\over 2c^2r^2}~, \eqno(4.30) 
$$
\no where $m$ and $Q$ are mass and charge.  
The terms in ${1\over r}$ and ${1\over r^2}$ agree with the
Reissner-Nordstrom result for a point source.  For the present case of an
extended source, 
there will be additional terms of higher order in
${1\over r}$.
\vskip.5cm

\no (d) Charged Scalars.

In order to extract similar asymptotic relations from (3.17), it is
sufficient to assume
$$
V(R^2) = {B\over 2} R^2 + \ldots ~. \eqno(4.31)
$$
\no Then at large $r$ one has
$$
R^{\prime\prime} + {2\over r} R^\prime + {\nu^\prime-\lambda^\prime\over 2}
R^\prime + \biggl({\omega\over c}+{e\varphi\over\hbar c}\biggr)^2
e^{\lambda-\nu} R-e^\lambda BR = 0~. \eqno(4.32)
$$
\no Let us carry $R$ to the same order as $\lambda$ in ${1\over r}$.
Therefore, let us make the ansatz:
$$
R = {e^{-\alpha r}\over r}\biggl(1 + {p_1\over r}\biggr)~. \eqno(4.33)
$$
\no Then
$$
R^{\prime\prime} + {2\over r} R^\prime =
\biggl(\alpha^2 + {2p_1\alpha\over r^2}\biggr) R~. \eqno(4.34)
$$
\no Eq. (4.32) may now be written to order ${1\over r^2}R$ as follows:
$$
\eqalign{\biggl(\alpha^2 + {2p_1\alpha\over r^2}\biggr) &-
{2m\alpha\over r^2} + \biggl(1+{4m\over r}+{2\ell_2+8m^2\over r^2}\biggr)
{\omega^2\over c^2}
+ {2e\over\hbar c^2}Q\omega\biggl({1\over r}+{4m\over r^2}\biggr)\cr
&+{2e\omega q_2\over\hbar c^2r^2}+{e^2Q^2\over\hbar^2c^2r^2}-
\biggl(1+{2m\over r}+{\ell_2+2m^2\over r^2}\biggr)B = 0~. \cr} \eqno(4.35)
$$
\no From (4.35) one finds
$$
\eqalignno{&B = \alpha^2+{\omega^2\over c^2}\geq 0 & (4.36) \cr
&B = {2\omega^2\over c^2}+{e\over\hbar c^2}{Q\omega\over m} & (4.37) \cr
&(p_1-m)\alpha + 2m^2\alpha^2+g(Q) = 0 & (4.38) \cr}
$$
\no where
$$
g(Q) = {e^2\over 2\hbar^2c^2}Q^2-{K\omega\over 2m}{e\over\hbar c^2}
Q^3+{e\omega\over c}q_2~. \eqno(4.39)
$$
\no By (4.36) and (4.37) the tail of $R(r)$ is determined by
$$
\alpha^2 = {\omega^2\over c^2}+{\ell\omega\over\hbar c^2m} Q~. \eqno(4.40) 
$$
\no If the mass is very large
$$
\alpha \cong {\omega\over c}~. \eqno(4.41)
$$
\ve

\ce {V.  THE SPECIAL RELATIVISTIC EIGENVALUE PROBLEM}
\vskip.5cm

The general relativistic  
equations that must be satisfied are (3.11)-(3.18) subject to
the boundary conditions (3.19)-(3.22) as well as (3.33) and (3.34),
after the dilaton equations have been deleted. 

The qualitative nature of the solutions to these equations is already
determined by the eigenvalue problem generated by the special relativistic
limit of (3.17), namely
$$
R^{\prime\prime} + {2\over r}R^\prime + {\omega^2\over c^2} R-
{\partial V\over\partial R} = 0 \eqno(5.1)
$$
\no which holds in the limit at large distances.

To discuss (5.1) introduce the auxiliary function:
$$
H = {1\over 2}(R^\prime)^2-V(R^2)+{1\over 2}{\omega^2\over c^2}R^2~. \eqno(5.2)
$$
\no By (5.1)
$$
{dH\over dr} = -{2\over r}(R^\prime)^2\leq 0~. \eqno(5.3)
$$
\no It follows that the representative point in the phase plane (the
$RR^\prime$-plane) will always
move towards lower values of $H$ unless $R^\prime=0$.
Hence, if
$H(R,R^\prime)$ is pictured as a surface above the $RR^\prime$-plane, the
representative point will fall toward minima on this surface.

The extrema in the $R^\prime$-direction are given by
$$
{\partial H\over\partial R^\prime} = R^\prime=0~. \eqno(5.4)
$$
\no This locus is the $R$-axis.  It is a minimum in the $R^\prime$
direction since
$$
{\partial^2H\over\partial(R^\prime)^2} = 1>0~. \eqno(5.5)
$$
\no The extrema in the $R$-direction depend on $V(R^2)$.

For definiteness assume that $V(R^2)$ may be represented by the
polynomial
$$
V(R^2) = {B\over 2} R^2+{C\over 4}R^4 + {D\over 6}R^6 \eqno(5.6)
$$
\no where
$$
B>0, \quad C<0, \quad D>0~. \eqno(5.7)
$$
\no Since $D$ is positive, the field energy is bounded below.  Then
$$
\eqalignno{{\partial H\over\partial R}~ &= -{\partial V\over\partial R}
+{\omega^2\over c^2} R = -(BR+CR^3+DR^5) + {\omega^2\over c^2}R & (5.8) \cr
{\partial^2H\over\partial R^2} &= -{\partial^2V\over\partial R^2}+
{\omega^2\over c^2} = -(B+3CR^2+5DR^4) + {\omega^2\over c^2}~.& (5.9) \cr}
$$
\no By (5.8) the extrema in the $R$-direction lie at the roots of
$$
{\partial H\over\partial R} = (\tilde B+CR^2+DR^4)R = 0 \eqno(5.10)
$$
\no where
$$
\tilde B=B-{\omega^2\over c^2}~. \eqno(5.11)
$$
\no The five roots of (5.10) are
$$
R=0 \eqno(5.12)
$$
\no and
$$
\eqalignno{R_{++} &= +\biggl\{{|C|+E\over 2D}\biggr\}^{1/2} \qquad
R_{-+} = -R_{++} & (5.13) \cr
R_{+-} &= +\biggl\{{|C|-E\over 2D}\biggr\}^{1/2} \qquad
R_{--} = -R_{+-} & (5.14) \cr}
$$
\no Here
$$
E^2 = C^2-4D\tilde B~. \eqno(5.15)
$$
\no Then
$$
-\biggl({\partial^2H\over\partial R^2}\biggr)_{R=0} = \tilde B~. \eqno(5.16)
$$
\no We now require $\tilde B = B-\omega^2/c^{2} > 0$.  Then the origin is a maximum
in the $R$-direction.

At $R_{++}$ and $R_{+-}$ we have
$$
\eqalignno{\biggl({\partial^2H\over\partial R^2}\biggr)_{++} &=
-{E\over D}(|C|+E) < 0 & (5.17) \cr
\biggl({\partial^2H\over\partial R^2}\biggr)_{+-} &=
{E\over D} (|C|-E) > 0 & (5.18) \cr}
$$
\no since $|C| > E$.

Hence the points $(R_{+-},0)$ and $(-R_{+-},0)$ are minima in both $R$ and
$R^\prime$ directions.
The point $(0,0)$ is a minimum in the $R^\prime$ direction but a maximum
in the $R$-direction and is therefore a saddle point.
Similarly the points $(R_{++},0)$ and $(-R_{++},0)$ are saddle points.

The phase portrait of $H(R,R^\prime)$ is shown in Fig. 1.
We shall assume in this figure that 

$$
H(R_{++},0) > H(0,0) \quad \hbox{or} \quad H(R_{++},0) > 0 \eqno(5.19)
$$
\no To satisfy this condition, $D$ is restricted by the following inequality:
$$
{3\over 16} {C^2\over\tilde B} \geq D~. \eqno(5.20)
$$

If $V(R)$ were a higher order polynomial or some other functional, the
phase portrait could have more structure than is shown in the figure.  It
will become clear, however, that only that portion of the surface near the
origin is relevant here.

Let $C_0$ and $C_{++}$ be iso-$H$ curves passing through (0,0) and
$(R_{++},0)$ .  Require that $D$ be restricted by (5.20).  Then if the
initial point of the solution curve lies anywhere in the phase plane between
$C_0$ and $C_{++}$, this curve must terminate, according to (5.3), at (0,0),
$(R_{+-},0)$ or $(-R_{+-},0)$.  In the soliton problem the initial point
of the solution curve must lie on the $R$-axis, and if it also lies between
$C_0$ and $R_{++}$ it must terminate at either one of the two attractors,
$R_{+-}$ and $-R_{+-}$, or at the origin.

In Fig. 2 the region between $C_0$ and $C_{++}$ is expanded.
The solution curves are shown starting at $p_0,~s_0$, and $q_0$ and
ending at $P,S$, and $Q$.  In Fig. 3 the same solutions are shown in 
configuration space.

Every solution curve starting on the axis between $R_{+-}$ and 
$R_{++}$ and, depending on its initial point, lies in one of two classes
going to either the left $(P)$ or right $(S)$ attractor, unless it goes
to the origin.  The relation between curves in the $p$ and $s$ classes
and the eigensolution is
$$
\lim_{p_o-s_o\to 0}\biggl[{R(p)+R(s)\over 2}\biggr] = R(q) \eqno(5.21)
$$
\no where $p,s$, and $q$ refer to the three curves going to $P,S$, and
$Q$.

By bringing $p_o$ and $s_o$ together one may determine the eigenvalue 
$q_o$ with arbitrary accuracy.

Similarly Eq. (5.21) is a prescription for obtaining the eigensolutions 
$R(q)$ with arbitrary accuracy.

Note that at large $r$ we have approximately
$$
\eqalignno{R(p) &= A+u(p) & (5.22) \cr
R(s) &= -A+u(s) & (5.23) \cr
R(q) &= u(q) & (5.24) \cr}
$$
\no where 
$A = R_{+-}$.  Since $(\pm A,0)$ are 
constant solutions of (5.1), they are also roots of the following polynomial
and hence extrema of $H$:
$$
\biggl({\partial H\over\partial R}\biggr)_{\bar R} =
{\omega^2\over c^{2}} \bar R-\biggl({\partial V\over\partial R}\biggr)_{\bar R} =
0~, \qquad \bar R = (\pm A,0)~. \eqno(5.25)
$$
\no Here $u_p,u_s$, and $u_q$ are spherical Bessel functions that solve
(5.1) after it is linearized near $\pm A$ and zero.  The linearized equation
is
$$
u^{\prime\prime} + {2\over r} u^\prime +
\biggl({\partial^2H\over\partial R^2}\biggr)_{\bar R} u = 0 \qquad
\bar R = (\pm A,0) \eqno(5.26)
$$
\no and
$$
\eqalignno{u(p) &= {a_p\over r}\sin(\tilde\omega_pr + \epsilon_p) & (5.27)\cr
u(s) &= {a_s\over r}\sin(\tilde\omega_sr + \epsilon_s) & (5.28)\cr
u(q) &= {a_q\over r} e^{-\alpha r} & (5.29) \cr}
$$
\no where
$$
\eqalignno{\tilde\omega_p^2 &= \tilde\omega_s^2 = \biggl({\partial^2H\over
\partial R^2}\biggr)_A & (5.30) \cr
\alpha^2 &= -\biggl({\partial^2H\over\partial R^2}\biggr)_0 & (5.31) \cr}
$$

The functions $u(p)$ and $u(s)$ describe the spiral motion toward the
attractors while $u(q)$ describes the exponential approach to the
saddle point.  Note that the oscillations of $u(p)$ and $u(s)$ begin to
cancel in the sum ${1\over 2} (R(p)+R(s))$ because of the phase
difference between the arguments of $u(p)$ and $u(s)$ 
that arise because
the motions around $S$ and $P$ are clockwise and counterclockwise
respectively $(\tilde\omega_p=-\tilde\omega_s)$.  
The phase difference for the actual numerical solutions
is shown in Fig. 4a. In the large distance
limit the oscillatory behavior disappears since ${1\over 2}(u(p)+u(s))$
must approach $u(q)$.

As $p_o$ and $s_o$ are brought closer to $q_o$ the oscillatory behavior of
$p$ and $s$ recedes to infinity and both $p$ and $s$ approach the eigensolution,
$q$, which terminates at the origin with no oscillations.
In Fig. (4b) this eigensolution is shown in configuration space.

The preceding discussion continues to hold as the initial values, $p$ and
$s$, are increased.  It may then happen that the $p$ and $s$ curves cross the
$R^\prime$ axis several times before being caught on one of the attractors.
In that case the eigensolutions will node a corresponding number of times
and the resultant field structure may be described as an excited state of the
soliton.

On the other hand, if the initial point lies inside the separatrix, the 
solution curve will always end at the nearby attractor.  According to one of
the boundary conditions that we have imposed, namely $R(\infty) = 0$, such
a curve is not an eigensolution.  If the boundary conditions were relaxed,
however, so that $R(\infty)$ is interpreted as a vacuum expectation value,
as in the case with gauge solitons, this interior curve could be regarded as
belonging to a continuum of eigenstates.

The preceding qualitative discussion is based entirely on equation (5.1)
but the illustrating figures are based on the numerical solution of the 
complete set of simultaneous differential equations. This is possible
because the  qualitative behavior of the solution to (5.1) is very similar
to the corresponding solutions of (3.17) viewed as one member of the 
set of  simultaneous differential equations.

\ve

\ce {VI.  THE GENERAL RELATIVISTIC EIGENVALUE PROBLEM}
\vskip.5cm

Although we have not found an $H$-functional appropriate to the full set of
equations, nevertheless the numerically obtained solutions of the original
equation (3.17) behave qualitatively in the way just described for the
simplified equation (5.1).

We may gain additional information by regarding the full equation (3.17)
as a perturbation on (5.1) and by confining the discussion to the
$RR^\prime$-plane.

Consider, for example, the region of phase space where 
$\nu=\lambda=\nu^\prime =\lambda^\prime \simeq 0$.   Then
$$
R^{\prime\prime} + {2\over r} R^\prime +
\biggl({\omega\over c} + {e\over \hbar c}\varphi\biggr)^2 R -
{\partial V\over\partial R} = 0 \eqno(6.1)
$$
\no and
$$
\eqalign{{dH\over dr} &= R^\prime\biggl[R^{\prime\prime}+\omega^2R -
{\partial V\over\partial R}\biggr] \cr
&= (R^\prime)^2\biggl\{-{2\over r}-\biggl[\biggl({\omega\over c}+
{e\over\hbar c}\varphi\biggr)^2 - {\omega^2\over c^2}\biggr]
{R\over R^\prime}\biggr\} ~. \cr} \eqno(6.2)
$$
\no For very large $R$
$$
{dH\over dr} \cong {2(R^\prime)^2\over r}
\biggl\{-1+{\omega e\over\hbar c^2}{Q\over \alpha}\biggr\} \eqno(6.3)
$$
\no if
$$
\varphi\cong {Q\over r} \quad \hbox{and} \quad
R\cong {e^{-\alpha r}\over r}~. \eqno(6.4)
$$
\no Then 
$$
\lim_{r\to\infty}{dH\over dr}=0~; \eqno(6.5)
$$
\no but it may be of either sign depending on the sign and magnitude of the
second term.  If the sign becomes positive after the representative point
crosses the separatrix, it will settle  
into a
closed orbit about either the left or right minimum, depending on whether
it is caught in the left or right lobe.  Then in
configuration space $R$ will not approach a constant at infinity but
will continue to oscillate.

We also note that the sign of $dH/dr$ may be switched 
outside the separatrix by slightly changing the magnitude of $\varphi$.
If this happens when the representative point has come close in the approach
to either lobe, it may happen that it finally terminates
in the opposite lobe.  In configuration space, this behavior will appear
as a discontinuous change in the solution curve $R(r)$ caused by a small
change in the parameter $\varphi(0)$.

None of these remarks invalidate the exact, non-perturbative
procedure for determining the
eigenvalues and eigensolutions of the set (3.11)-(3.18) (always 
excluding the dilaton equation). 
The most important result of the
numerical work is the observation that the $R(r)$ solutions divide into two
classes when the four general relativistic differential equations are
simultaneously integrated.  This key feature of the special relativistic
problem is exactly preserved.  Now, however, associated with $R(p),~R(s)$
and $R(q)$ are the families $[\lambda(p),~\nu(p),~\varphi(p)],~
[\lambda(s),~\nu(s),~\varphi(s)]$ and $[\lambda(q),~\nu(q),~
\varphi(q)]$.  Then in addition to (5.21), we have
\vskip 0.5cm

$$
\lim_{p_0-s_0\to 0} 
\biggl[{\lambda(p) + \lambda(s)\over 2}\biggr] = \lambda(q) \eqno(6.6)
$$

\vskip 0.5cm
\no with similar relations for $\nu$ and $\varphi$, so that in the
limit $[R(q),~\lambda(q),~
\nu(q),~\varphi(q)]$ is a complete set of solutions
corresponding to $q_0$.  
We also note that the $p$ and $s$ curves for $\lambda$, $\nu$ and $\varphi$
are nearly the same while the corresponding curves for $R$ are of
course quite distinct at large $r$.

There is also the following approximate procedure:
Carry the integration into the region where the gravitational potentials
become asymptotic.  Up to this point, say $\bar r$, which is somewhat
arbitrary, approximate the four functions by their $s$ and $p$ average:
${1\over 2}[R(p)+R(s)],~{1\over 2}[\lambda(p)+\lambda(s)] \ldots$.  
We shall choose as a criterion for determining $\bar r$ the condition:
$$
\lambda(\bar r) + \nu(\bar r) = 0~. \eqno(6.7)
$$
\no Beyond this point approximate the eigensolution $R(q)$ by its asymptotic
form (a spherical Bessel function).  Then match the asymptotic form of
$R(q)$ to ${1\over 2}[R(p)+R(s)]$ at $\bar r$ as follows:
$$
\eqalignno{{a_qe^{-\alpha\bar r}\over \bar r} &= {1\over 2}
\bigl(R(p) + R(s)\bigl)_{\bar r} & (6.8)\cr
\biggl({a_qe^{-\alpha r}\over r}\biggr)^\prime_{\bar r} &= {1\over 2}
\bigl(R^\prime(p) + R^\prime(s)\bigl)_{\bar r}~. & (6.9)\cr}
$$
\no To complete the integration beyond $\bar r$ feed the so obtained
$R(q)$ into the complete set of differential equations while taking
${1\over 2}\bigl[\lambda(p)+\lambda(s)\bigr]_{\bar r} \ldots$ as initial
values of the other functions.
If the point defined by (6.7) is difficult to reach, one may instead  
choose $\bar{r}$ as the point at which the curves $s$ and $p$
begin to separate. Figure 4b was obtained by this alternative procedure.

To find the functions $\varphi(r)$, $\lambda(r)$ and $\nu(r)$, corresponding 
to the eigenfunction $R(r)$, we have integrated backwards by feeding
$R(r)$ into the appropriate differential equations while imposing
initial conditions determined by the asymptotic forms (4.30) and
(4.31) and computed for a point $r_{\infty}$ and a mass $m$. The
values of $r_{\infty}$ and $m$ must be adjusted to agree with 
the boundary conditions at $r=0$. The results of this backward 
integration are presented in Fig. 5.

Additional checks may be obtained by integrating equations
(3.13) and (3.25) as follows:
$$
\eqalignno{\bigl[\nu(r)-\lambda(r)\bigr]^\infty_0 &= {1\over r}
\int^\infty_0 dr(e^\lambda-1) + K\int^\infty_0 drr 
[V(R^2)e^\lambda + 2 \Delta]  = J_1 &
(6.10) \cr
\bigl[\nu(r) + \lambda(r)\bigr]^\infty_0 &= -K \int^\infty_0 drr
\biggl[(R^\prime)^2 + 
\biggl({\omega\over c} +{e \over{\hbar c}} \varphi \biggr)^2  e^{\lambda-\nu}
{R^2\over c^2}\biggr] = J_2~. & (6.11) \cr}
$$
\no Now imposing the boundary conditions $\nu(\infty) = \lambda(\infty) = 0$
we have
$$
\eqalignno{\lambda(0) - \nu(0) &= J_1 & (6.12) \cr
\lambda(0) + \nu(0) &= -J_2 < 0 & (6.13) \cr}
$$

These equations may also be used iteratively to improve approximate
solutions already obtained.

The discrete spectrum of $R$-eigensolutions will induce discrete spectra
on the other fields.  As a consequence the mass of the soliton, which is
identified by the coefficient of the $1/r$ term in the asymptotic gravitational
potential, is also limited to a discrete spectrum.

It has been
noted that a small change in $\varphi(0)$, like a small change in
$R(0)$, may cause the solution curve in the $RR^\prime$-plane to jump
between the $P$ and $S$ lobes.  Therefore we may divide the
$[\varphi(0),~R(0)]$ plane into + and - areas according to whether the
solution curve terminates in the $P$ or $S$ lobe.  More generally, one
may divide the 4-dimensional space of initial values with coordinates
$[\lambda(0),~\nu(0),~\varphi(0),~R(0)]$ into + and - subspaces.
Then any boundary between a + and - subspace will describe a hypersurface of
eigenvalues.  We have investigated only a limited region of the 4-dimensional
parameter space.
\ve

\ce{VII. DISCUSSION}
\vskip.5cm

General relativistic solitons have also appeared in the literature on boson
stars.$^5$  The purpose and methods of this paper are quite different, however,
since they are more directly related to the first reference $^1$ and to the
possible existence of non-topological solitons in fundamental completions of
Einstein theory.  The main result of the present work is the demonstration
that the method of analysis employed in the earlier work may be extended to
many coupled fields.  While the field structures studied in this paper couple
the gravitational, electromagnetic, nonlinear scalar, and dilaton fields,
there are many other interesting possibilities that are tractable by the same
method, such as gauge solitons and other structures generated by
Kaluza-Klein-like theories.

Since there has been a particular emphasis in our investigation on the
nonlinear scalar potential, the generality of this potential is of interest.
Although a particular polynomial representation was chosen, it is clear
that the analysis depends only on the distribution of minima and saddle
points of the dissipative function along the field-axis in phase space and
therefore on the distribution of maxima and minima in the potential.  One also
notes that there are essentially only two kinds of potential, namely Higgs
and inverted Higgs, or volcano, which are related by the interchange of minima
with saddle points.  If the potential is of the volcano type, there is a 
saddle point in the dissipative function flanked by two minima and if it is
of the Higgs form, there is a central minimum bracketed by two saddle points.
If this pattern of three extrema is centered at $r=0$, then the eigenstates
of the former 
vanish at great distances.  In the Higgs case there is a continuum of 
eigensolutions and in the case of the volcano potential there is a discrete
set of eigensolutions, as we have discussed.  If these patterns are not
centered at $R=0$, however, then the corresponding eigenfunctions do not
vanish at infinity.  In this case one has the situation encountered in gauge
theories where the limiting value at infinity is interpreted as a vacuum
expectation value.  If the potential is general with several minima, there
should be both continuous and discrete spectra associated with it.  
For example, the potential may be Higgs-like at weak fields and
volcano-like at strong fields.
If the
boundary conditions are interpreted in terms of vacuum expectation values, 
then the physical constants appearing in the field Lagrangian are encoded as
vacuum expectation values.

Since our remarks concerning the dilaton field may also be relevant
to certain string models, let us consider two examples.$^{(6,7)}$

A dimensionally reduced superstring in four dimensions can be described 
in terms of N=4 supergravity: the action associated with the SO(4)
version of this theory is 
$$
I(SO(4)) = \int d^{4}x \left[ -R + 2 \partial^{\mu}\phi \partial_{\mu}\phi 
- (e^{-2 \phi} F_{\mu\nu}F^{\mu\nu} +
e^{2 \phi} \tilde{G}_{\mu\nu} \tilde{G}^{\mu\nu} )\right] \eqno(7.1)
$$
\no as expressed for example in reference (6). Dropping the $G$ term one
sees that the dilaton field $\phi$ appears here in the same way as in
our equation (1.2). Therefore we may conclude that the same conclusion
also holds here, namely, that non-topological solitons do not exist
for this theory.

As a second example consider the 5-brane soliton that is a solution
of the field equations describing the low energy heterotic 
string.$^{(7)}$ This soliton embodies a 4-dimensional Yang-Mills instanton
coupled to dilaton and axion fields according to superstring theory.
The dilaton equation in this case is$^{(7)}$
$$
\nabla \phi = \pm {1\over 120} \alpha^{\prime} 
\epsilon^{\mu \nu \rho \sigma} Tr F_{\mu \nu} F_{\rho \sigma} \eqno(7.2)
$$

This equation also resembles our equation (3.4) but with the following
differences: the source term in (7.2) is pseudoscalar rather than
scalar and moreover it lies in a gauge algebra; the instanton
exists not in  Minkowski but in Euclidean space, and the dilaton at
spatial infinity does not vanish but approaches its vacuum expectation
value. The field of this topological soliton is everywhere finite
as required and its extension is determined by the size of the 
instanton. The full structure has not been presented, however, and its
interpretation in conventional spacetime depends on how the 5-branes
wrap around topologically
non-trivial surfaces in the internal compactification manifold.
Since this topological structure is is therefore essentially different
from the non-topological four dimensional solution studied here, 
our remarks about the dilaton field do not apply to this example
although they do to the preceding case. 
On the other hand, the procedure described here may be useful for 
investigating non-abelian as well as abelian theories, and therefore
for describing topological as well as non-topological solitons.

\ve

\ce {REFERENCES}
\vskip.5cm

\item{1.} R. Finkelstein, R. Lelevier, and M. Ruderman, Phys. Rev. {\bf 83},
326 (1951).
\item{2.} R. Friedberg, T. D. Lee, and A. Sirlin, Phys. Rev. D{\bf 13},
2739 (1996).
\item{3.} R. Friedberg, T. D. Lee, and Y. Pang, Phys. Rev. D{\bf 35},
3640, 3698 (1981).
\item{4.} A. C. Cadavid and R. J. Finkelstein, 
Phys. Rev. D{\bf 57}, 7318 (1998).

\item{5.} P. Jetzer, Phys. Rep. {\bf 220}, 163 (1992).
\item{6.} 
R. Kallosh, A. Linde, T. Ortin, A. Peet and A. Van Proyen, 
Phys. Rev. D{\bf 46}, 5278 (1992).
\item{7.} A. Strominger, Nucl. Phys. B{\bf 343}, 167 (1990).

\ve 

\ce {FIGURE CAPTIONS}

\vskip 0.5cm
\no
Fig. 1. Phase portrait of $H(R,R')$. $Q$, $Q'$ and $Q''$ are saddle
points. $S$ and $P$ are minima. The $R$ coordinates of $Q$, $Q'$ and $Q''$
are $R_{-+}$, $0$ and $R_{++}$ respectively. The $R$ coordinates of
$S$ and $P$ are $R_{+-}$ and $R_{--}$ respectively.

\vskip 0.5cm
\no
Fig. 2. Solution curves $s$ and $p$ bounding the eigensolution
$q$, shown in the phase plane. 

\vskip 0.5cm
\no
Fig. 3. Solution curves $s$ and $p$ bounding the eigensolution
$q$, shown in configuration space.

\vskip 0.5cm
\no
Fig. 4. (a) The asymptotic behavior of the solutions $s$ and $p$,
illustrating the approximate phase cancellation at large $r$.
(b) An eigensolution of $R(r)$ in configuration space as determined 
by the numerical procedure described in the text.

\vskip 0.5cm
\no
Fig. 5. (a) The gravitational eigensolutions $\lambda(r)$ and $\nu(r)$.
(b) The electromagnetic  eigensolution $\varphi(r)$.

\ve
\centerline{\epsfxsize 8.0 truein \epsfbox{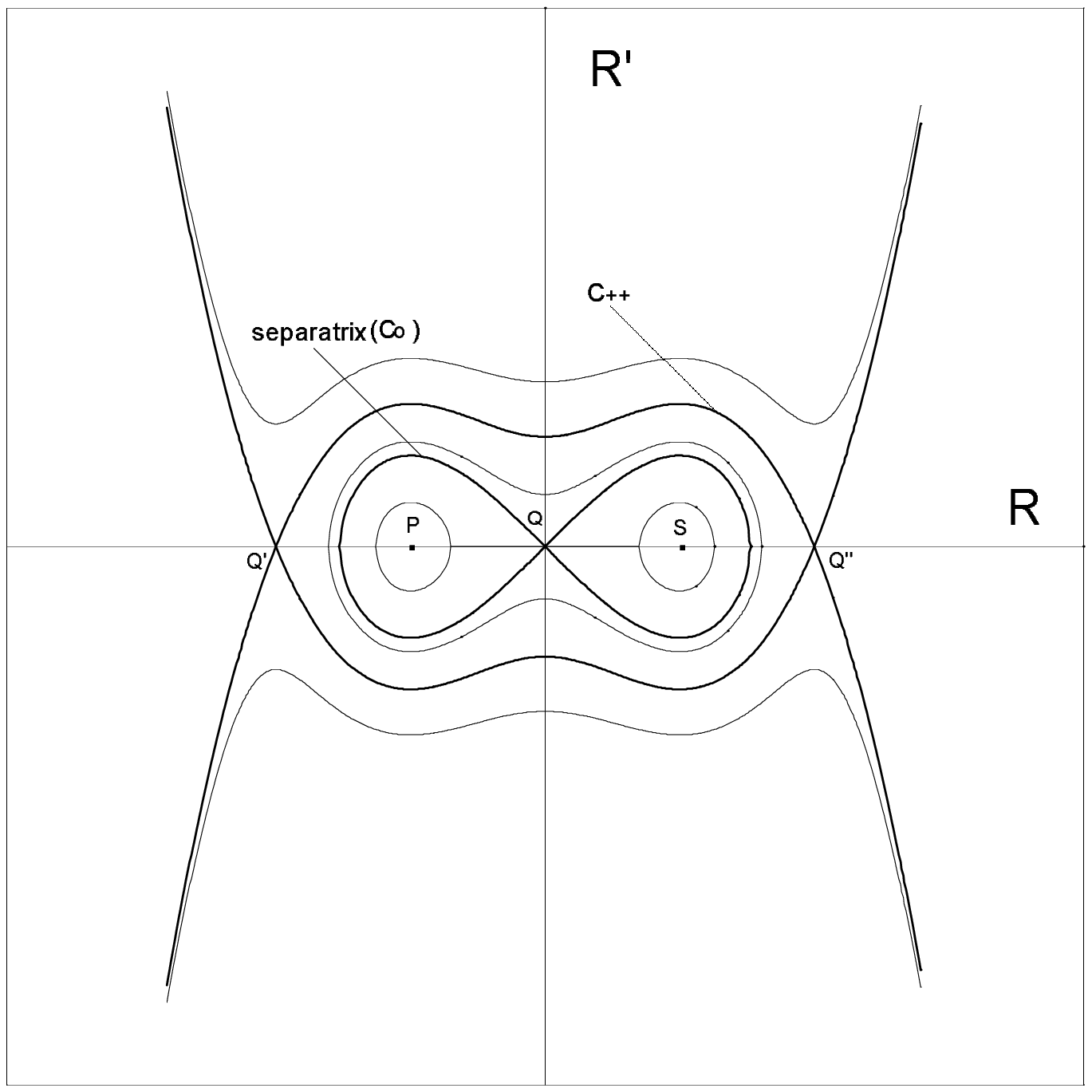}} 
\ve
\centerline{\epsfxsize 8.0 truein \epsfbox{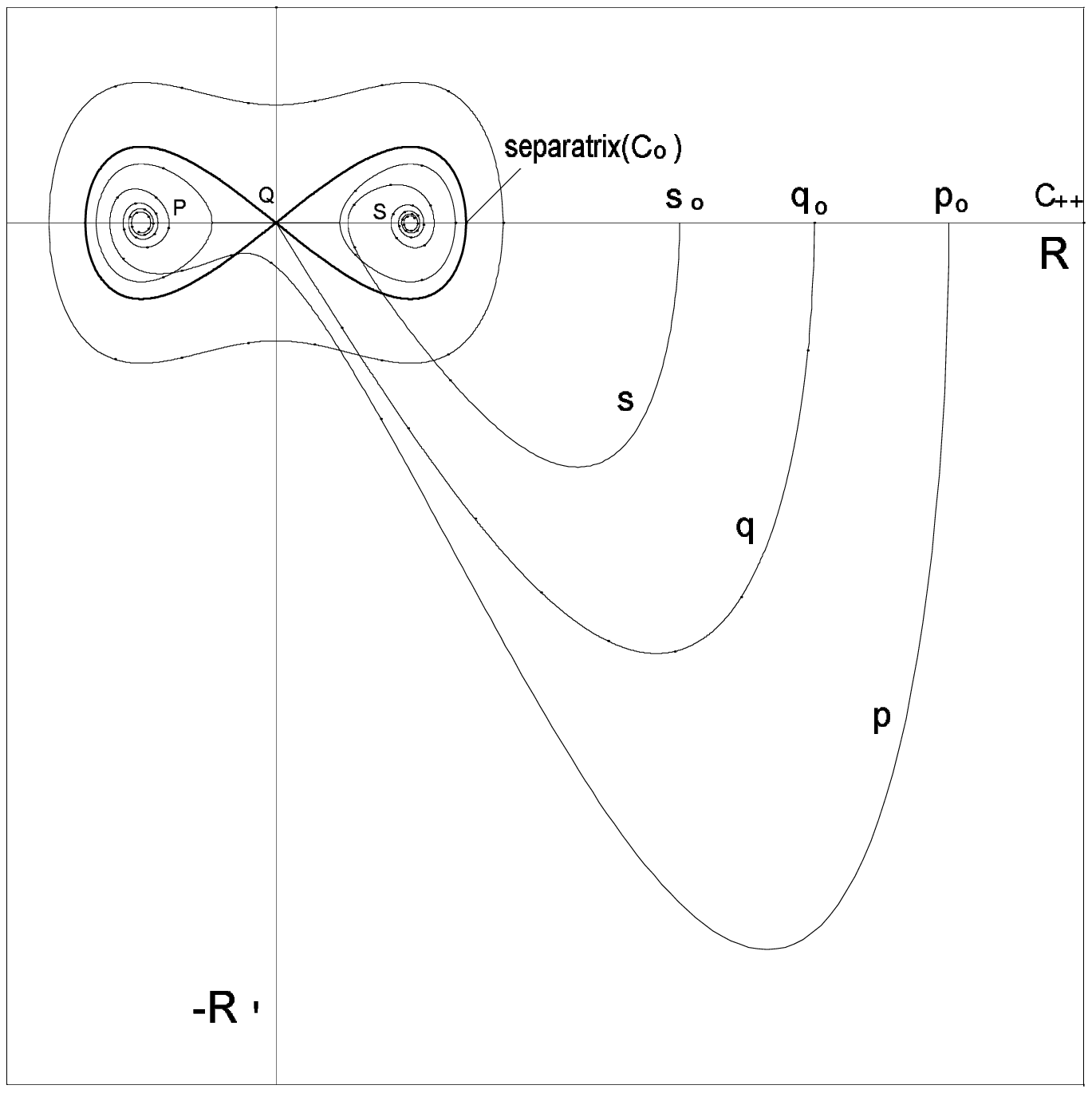}} 
\ve
\centerline{\epsfxsize 8.0 truein \epsfbox{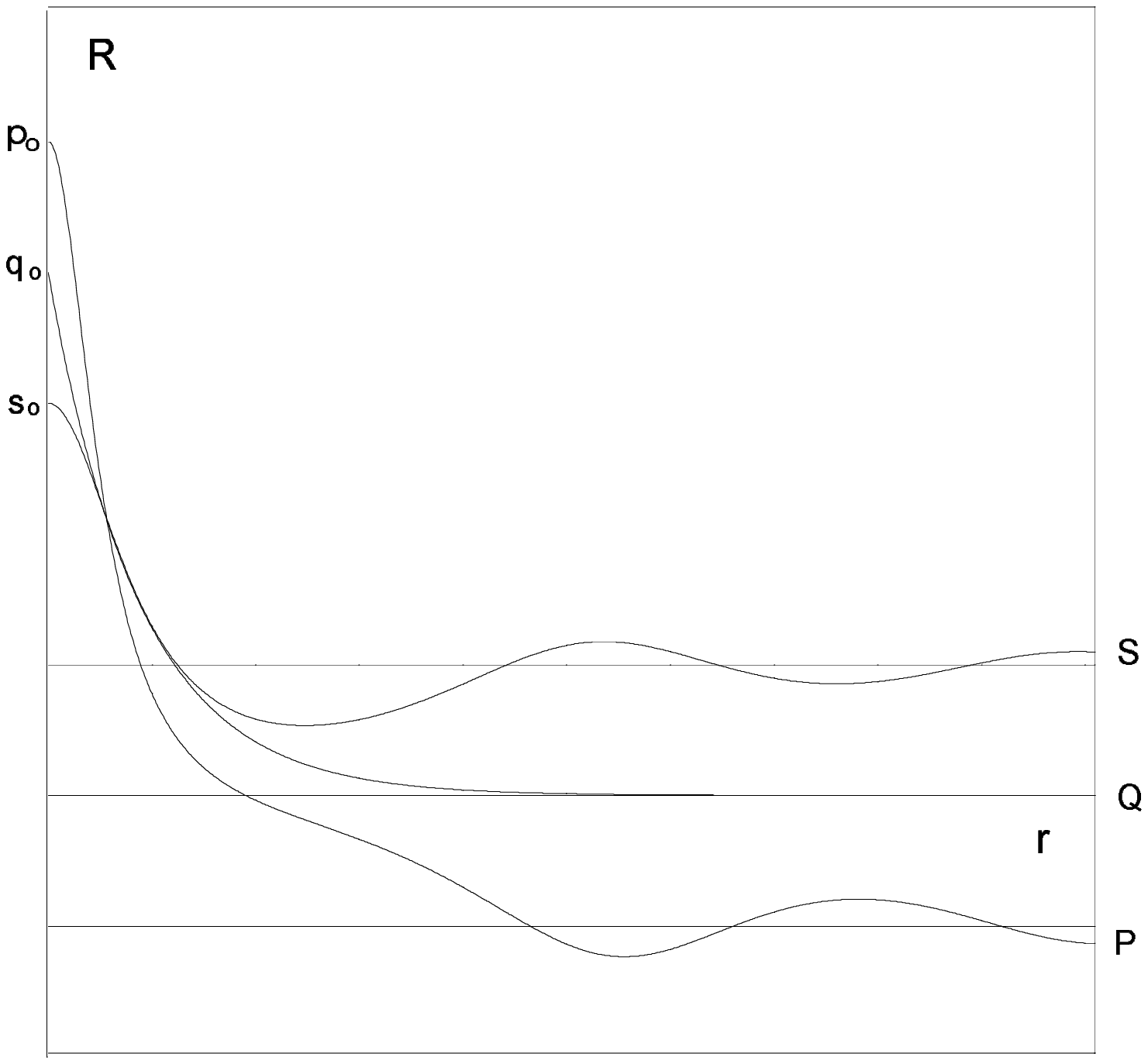}} 
\ve
\centerline{\epsfxsize 8.0 truein \epsfbox{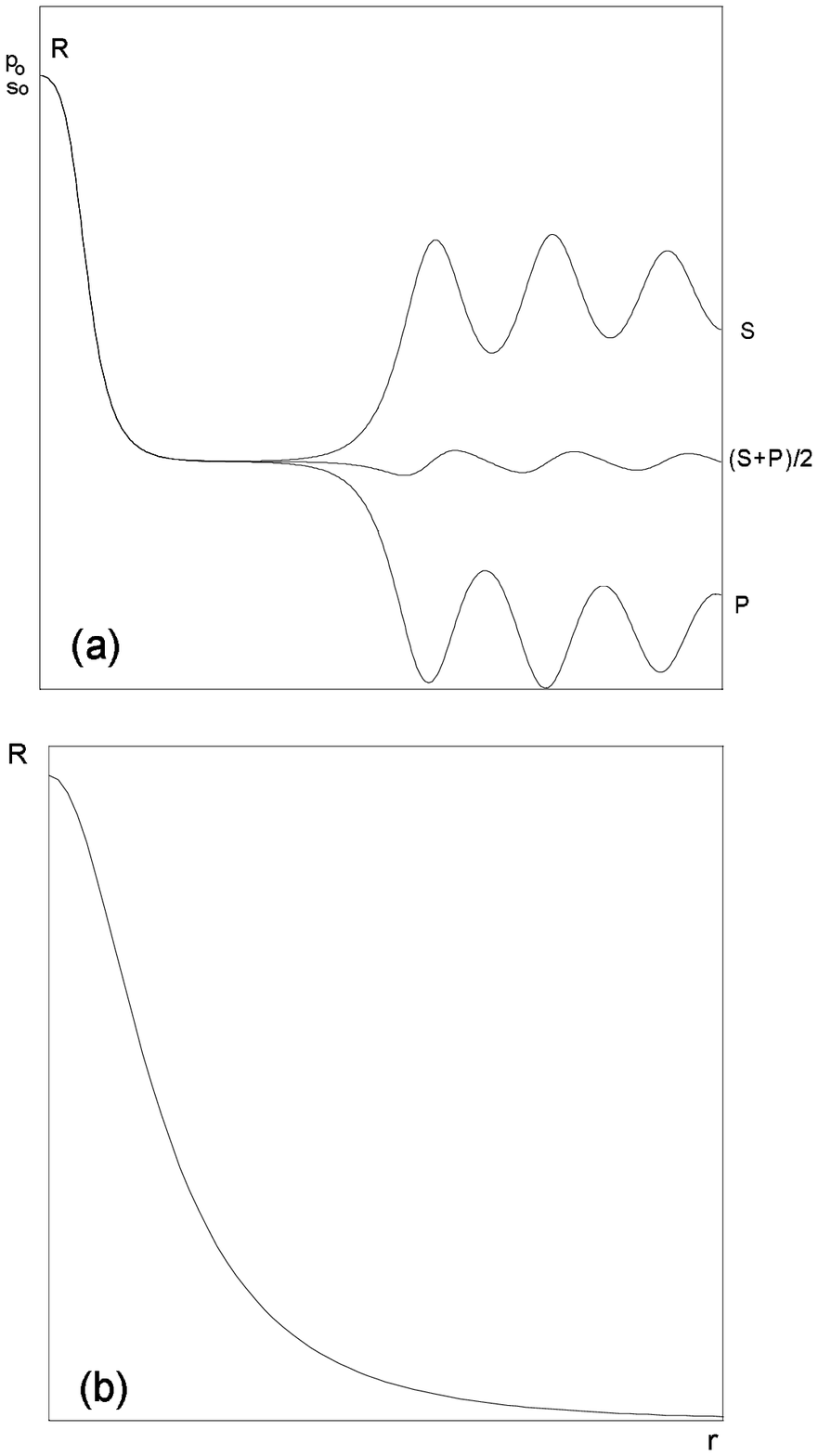}} 
\ve
\centerline{\epsfxsize 8.0 truein \epsfbox{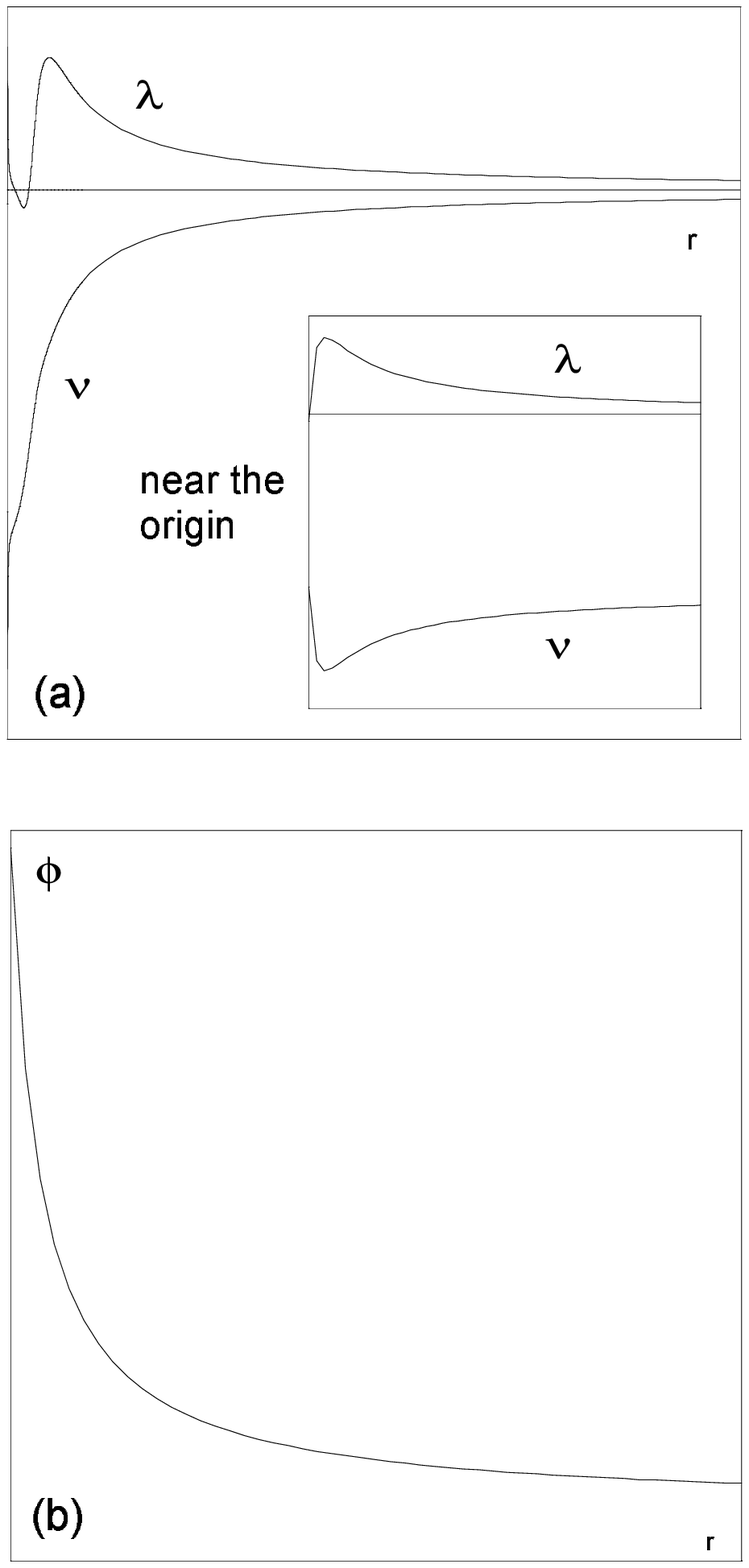}} 
\ve

\end

\bye